\documentclass[reqno]{article}
\usepackage{amssymb,amsmath,cite,delarray,epsfig,hhline}
\newcommand{\lsim} {\buildrel < \over {_\sim}}
\newcommand{\gsim} {\buildrel > \over {_\sim}}

\begin{document}
\begin{flushright}
\begin{tabular}{l}
OSU-HEP-05-03
\end{tabular}
\end{flushright}

\begin{center}
{\LARGE \bf Lepton Flavor Violation within a realistic SO(10)/G(224) Framework} \\
\vskip.3in ${\large\bf
K.S. ~Babu^a,\ Jogesh\ C.\ Pati^b,\ Parul\ Rastogi^b}$\\[0.1in]
$^a Oklahoma\ Center\ for\ High\ Energy\ Physics,\ Department\ of\
Physics$ \\[-0.03in]
$Oklahoma\ State\ University,\ Stillwater,\ OK\ 74078,\
USA$\\[0.1in]
${\large ^b Department\ of\ Physics,\ University\ of\ Maryland,\ College\ Park\  MD\ 20742,\ USA}$\\
February 16, 2005.
\end{center}

\vskip.25in
\begin{abstract}
Lepton flavor violation (LFV) is studied within a realistic
unified framework, based on supersymmetric SO(10) or an effective
G(224) = $SU(2)_L\times SU(2)_R\times SU(4)^c$ symmetry, that
successfully describes (i) fermion masses and mixings, (ii)
neutrino oscillations, as well as (iii) CP violation. LFV emerges
as an important prediction of this framework, bringing no new
parameters, barring the few SUSY parameters, which are assumed to
be flavor-universal at $M^*\gsim M_{GUT}$. We study LFV (i.e.
$\mu\to e\gamma,~\tau\to\mu\gamma,~\tau\to e\gamma$ and $\mu N\to
e N$) within this framework by including contributions {\it both}
from the presence of the right handed neutrinos as well as those
arising from renormalization group running in the post-GUT regime
($M^*\to M_{GUT}$). Typically the latter, though commonly omitted
in the literature, is found to dominate. Our predicted rates for
$\mu\to e\gamma$ show that while some choices of ($m_o,~m_{1/2}$)
are clearly excluded by the current empirical limit, this decay
should be seen with an improvement of the current sensitivity by a
factor of 10--100, even if sleptons are moderately heavy ($\lsim$
800 GeV, say). For the same reason, $\mu-e$ conversion ($\mu N\to
e N$) should show in the planned MECO experiment. Implications of
WMAP and $(g-2)_{\mu}$-measurements are noted, as also the
significance of the measurement of parity-odd asymmetry in the
decay of polarized $\mu^+$ into $e^+\gamma$.
\end{abstract}
\newpage
{\large

\section{Introduction}

Individual lepton numbers (L$_e$, L$_\mu$ and L$_\tau$) being
symmetries of the standard model (SM)(with m$_\nu^i$ = 0),
processes like $\mu\to e \gamma$, $\tau \to \mu \gamma$ and $\tau
\to e \gamma$ are forbidden within this model. Even within simple
extensions of the SM (that permit m$_\nu^i \ne 0$), they are too
strongly suppressed to be observable. Experimental searches have
put upper limits on the branching ratios of these processes:
Br$\left(\mu\to e\gamma\right)\le 1.2\times 10^{-11} $\cite{mue},
Br$\left(\tau\to \mu\gamma\right)\le 3.1\times 10^{-7}
$\cite{taumu} and Br$\left(\tau\to e\gamma\right)\le 3.7\times
10^{-7}$\cite{taue}. The extreme smallness of these branching
ratios poses a challenge
 for physics beyond the standard model, especially for supersymmetric grand unified (SUSY GUT) models, as these generically possess new sources of lepton flavor violation that could easily lead to rates even surpassing the current limits.

In this paper, we study how lepton flavor violation (LFV) gets
linked with fermion masses, neutrino oscillations and CP violation
within a predictive SUSY grand unified framework, based on either
SO(10) \cite{SO(10)}, or an effective (presumably string derived)
symmetry G(224) = $SU(2)_L\times SU(2)_R\times SU(4)^c$ \cite{PS}.
The desirability of having an effective symmetry that possesses
SU(4)-color \cite{PS}, in view of the observed neutrino
oscillations and the likely need for leptogenesis as a means for
baryogenesis \cite{Yanagida,PatiLepto}, has been stressed
elsewhere (see e.g. \cite{JCPKeK}).

A predictive framework based on supersymmetric SO(10) or
G(224)-symmetry has been proposed by Babu, Pati and Wilczek (BPW)
in \cite{BPW}, which successfully describes the masses and mixings
of all fermions including neutrinos. In particular it makes seven
predictions, all in good accord with observations. This framework
was recently extended by us in Ref. \cite{BPR} to describe the
observed CP and flavor violations by allowing for phases in the
fermion mass matrices. Remarkably enough, this extension could
successfully describe the masses of all the quarks and leptons
(especially of the two heavier families), the CKM elements, the
observed CP and flavor violations in the K$^{\circ}-\overline{{\rm
K}^{\circ}}$ system (yielding correctly $\Delta$m$_K$ and
$\epsilon_K$) and the B$_d^{\circ}-\overline{{\rm B}_d^{\circ}}$
system (yielding the correct values of $\Delta$m$_{B_d}$ and
S$_{\psi K_S}$).

In this paper, we study lepton flavor violating processes, i.e.
$\mu\to e \gamma$, $\tau \to \mu\gamma$, $\tau \to e\gamma$ and
$\mu N\to e N$, within the same framework \cite{BPW,BPR}. The
subject of LFV has been discussed widely in the literature within
supersymmetric extensions of the standard model. (For earlier
works see Ref.~\cite{Masiero,BHS,HNTY}). Our work based on SUSY
SO(10) or G(224) differs from those based on either MSSM with
right-handed neutrinos (RHN's) \cite{Masiero,HNTY,MSSMRHN} or SUSY
SU(5) \cite{SU(5)RHN}, because for these latter cases the RHN's
are singlets and thereby their Yukawa couplings are {\it a priori}
arbitrary. By contrast, for G(224) or SO(10) the corresponding
Yukawa couplings are determined in terms of those of the quarks at
the GUT-scale (such as h($\nu^\tau$)$_{Dirac}\approx$ h$_{top}$)
(see Ref. \cite{BPW}). Thus the SUSY G(224)/SO(10)-framework is
naturally more predictive than the MSSM or SUSY SU(5)-framework.

In addition, our work differs from all others, including those
based on SUSY SO(10) \cite{SO(10)LFV} as well, in two other
important respects: First, we work within a predictive and
realistic framework \cite{BPW,BPR} which (as mentioned above)
successfully describes a set of phenomena -- i.e. (a) fermion
masses, (b) CKM mixings, (c) neutrino oscillations, (d) observed
CP and flavor violations in the K and B systems, as well as (e)
baryogenesis via leptogenesis \cite{JCPKeK}. As we will see,
lepton flavor violation emerges as an important prediction of this
framework, bringing no new parameters (barring the few
flavor-universal SUSY-parameters).

Second, we do a comprehensive study of LFV processes by including
contributions from three different sources: (i) the sfermion
mass-insertions, $\hat{\delta}^{ij}_{LL,RR}$, arising from
renormalization group (RG) running from M$^{*}$ to M$_{GUT}\sim
2\times 10^{16}$ GeV (where M$^{*}$ denotes the presumed
messenger-scale, with M$_{GUT}< $  M$^{*}\le$ M$_{string}$, at
which flavor-universal soft SUSY breaking is transmitted to the
squarks and sleptons, like in a mSUGRA model \cite{mSUGRA}), (ii)
the mass-insertions $(\delta^{ij}_{LL})^{RHN}$ arising from RG
running from M$_{GUT}$ to the right handed neutrino mass scales
M$_{R_i}$, and (iii) the chirality-flipping mass-insertions
$\delta^{ij}_{LR,RL}$ arising from $A-$terms that are induced
solely through RG running from M$^{*}$ to M$_{GUT}$ involving
gauginos in the loop. All the three types of mass-insertions:
$\hat{\delta}^{ij}_{LL,RR}$, $(\delta^{ij}_{LL})^{RHN}$ and
$\delta^{ij}_{LR,RL}$ are in fact fully determined in our model.
(See \cite{BPR} for details). Most previous works in this regard
have included only the second contribution associated with the RH
neutrinos in their analysis.\footnote{Barbieri, Hall and Strumia
(in Ref. \cite{BHS}) have discussed the relevance of the
contributions from the mass-insertions $\hat{\delta}^{ij}_{LL,RR}$
and those from the induced $A-$terms, but without a realistic
framework for light fermion masses and neutrino oscillations.} We
find, however, that it is the first and the third contributions
associated with post-GUT physics that typically dominate over the
second in a SUSY unified framework.

A brief review of our previous work and our results are presented
in the following sections.

\section{A brief review of the BPW framework and its extension}

The Dirac mass matrices of the sectors $u, d, l$ and $\nu$
proposed in Ref. \cite{BPW} in the context of SO(10) or
G(224)-symmetry have the following structure:

\begin{eqnarray}
\label{eq:mat}
\begin{array}{cc}
M_u=\left[
\begin{array}{ccc}
0&\epsilon'&0\\-\epsilon'&\zeta_{22}^u&\sigma+\epsilon\\0&\sigma-\epsilon&1
\end{array}\right]{\cal M}_u^0;&
M_d=\left[
\begin{array}{ccc}
0&\eta'+\epsilon'&0\\
\eta'-\epsilon'&\zeta_{22}^d&\eta+\epsilon\\0& \eta-\epsilon&1
\end{array}\right]{\cal M}_d^0\\
&\\
M_\nu^D=\left[
\begin{array}{ccc}
0&-3\epsilon'&0\\3\epsilon'&\zeta_{22}^u&\sigma-3\epsilon\\
0&\sigma+3\epsilon&1\end{array}\right]{\cal M}_u^0;& M_l=\left[
\begin{array}{ccc}
0&\eta'-3\epsilon'&0\\
\eta'+3\epsilon'&\zeta_{22}^d&\eta-3\epsilon\\0& \eta+3\epsilon&1
\end{array}\right]{\cal M}_d^0\\
\end{array}
\end{eqnarray}

These matrices are defined in the gauge basis and are multiplied
by $\bar\Psi_L$ on left and $\Psi_R$ on right. For instance, the
row and column indices of $M_u$ are given by $(\bar u_L, \bar c_L,
\bar t_L)$ and $(u_R, c_R, t_R)$ respectively. These matrices have
a hierarchical structure which can be attributed to a presumed
U(1)-flavor symmetry (see e.g. \cite{JCPKeK,BPR}), so that in
magnitudes 1 $\gg
\sigma\sim\eta\sim\epsilon\gg\zeta_{22}^u\sim\zeta_{22}^d\gg\eta'>\epsilon'$.
The entries $\epsilon$ and $\epsilon'$ are proportional to $B-L$
and are antisymmetric in family space (see below). Thus
($\epsilon, \epsilon'$)$\to -3(\epsilon, \epsilon')$ as $q \to l$.
Following the constraints of SO(10) and the U(1)-flavor symmetry,
such a pattern of mass-matrices can be obtained using a minimal
Higgs system consisting of $\mathbf{45_{H}}, \mathbf{16_{H}},
\overline{\mathbf{16}}_{\mathbf{H}}, \mathbf{10_{H}}$ and a
singlet {\it S} of SO(10), which lead to effective couplings of
the form \cite{JCPKeK,BPR}:
\begin{eqnarray}
\label{eq:Yuk} {\cal L}_{\rm Yuk} &=& h_{33}{\bf 16}_3{\bf
16}_3{\bf 10}_H  + [ h_{23}{\bf 16}_2{\bf 16}_3{\bf
10}_H(S/M) \nonumber \\
&+& a_{23}{\bf 16}_2{\bf 16}_3{\bf 10}_H ({\bf
45}_H/M')(S/M)^p+g_{23}{\bf 16}_2{\bf 16}_3{\bf 16}_H^d ({\bf
16}_H/M'')(S/M)^q] \nonumber \\ &+& \left[h_{22}{\bf 16}_2{\bf
16}_2{\bf 10}_H(S/M)^2+g_{22}{\bf 16}_2{\bf 16}_2 {\bf
16}_H^d({\bf 16}_H/M'')(S/M)^{q+1} \right] \nonumber \\ &+&
\left[g_{12}{\bf 16}_1{\bf 16}_2 {\bf 16}_H^d({\bf
16}_H/M'')(S/M)^{q+2}+ a_{12}{\bf 16}_1{\bf 16}_2 {\bf 10}_H({\bf
45}_H/M')(S/M)^{p+2} \right]~.
\end{eqnarray}

The mass scales $M'$, $M''$ and $M$ are of order $M_{\rm string}$
or (possibly) of order $M_{GUT}$\cite{FN6}. Depending on whether
$M'(M'')\sim M_{\rm GUT}$ or $M_{\rm string}$ (see \cite{FN6}),
the exponent $p(q)$ is either one or zero \cite{FN8}. The VEVs of
$\langle{\bf 45}_H\rangle$ (which is along $B-L$), $\langle{\bf
16}_H\rangle=\langle{\bf\overline {16}}_H\rangle$ (along
$\langle\tilde{\nu}_{RH}\rangle$) and $\langle S \rangle$ are of
the GUT-scale, while those of $\langle{\bf 10}_H\rangle$ and
$\langle{\bf 16}_H^d\rangle$ are of the electroweak scale
\cite{BPW,FN7}. The combination $ \mathbf{10_{H}}.\mathbf{45_{H}}$
effectively acts like a $\mathbf{120}$ which is antisymmetric in
family space and is along $B-L$. The hierarchical pattern is
determined by the suppression of the couplings by appropriate
powers of $M_{GUT}$/($M$, $M' or M''$). For details on how  Eq.
(\ref{eq:Yuk}) emerges from Eq. (\ref{eq:mat}) see Refs.
\cite{JCPKeK,BPW,BPR}.

The right-handed neutrino masses arise from the effective
couplings of the form \cite{FN30}:
\begin{equation}
\label{eq:LMaj} \mathcal{L}_{\mathrm{Maj}} = f_{ij} \mathbf{16}_i
\mathbf{16}_j \overline{\mathbf{16}}_H \overline{\mathbf{16}}_H/M
\end{equation}

\noindent where the $f_{ij}$'s include appropriate powers of
$\langle S \rangle/M$. The hierarchical form of the Majorana
mass-matrix for the RH neutrinos is \cite{BPW}:
\begin{eqnarray}
\label{eq:MajMM} M_R^\nu=\left[
\begin{array}{ccc}
x & 0 & z \\
0 & 0 & y \\
z & y & 1
\end{array}
\right]M_R
\end{eqnarray}

Following flavor charge assignments (see \cite{JCPKeK}), we have
$1\gg y \gg z \gg x$. The magnitude of M$_{\rm R}$ is estimated
(with $f_{33}\approx 1,~\langle
\overline{\mathbf{16}}_H\rangle\approx 2\times 10^{16}$ GeV and
$M\approx M_{\rm st}\approx 4\times 10^{17}$ GeV) as
\cite{BPW,JCPKeK}: $M_R = f_{33} \langle \overline{\mathbf{16}}_H
\rangle^2/M \approx (10^{15}\mbox{ GeV})(1/2\mbox{--}2)$.

Thus the Majorana masses of the RH neutrinos are given
by\cite{JCPKeK,BPW}:
\begin{eqnarray}
\label{eq:MajM}
\begin{array}{l}
\ M_{3} \approx M_R\approx 10^{15}\mbox{ GeV
(1/2--1)},\quad M_{2} \approx  |y^2|M_{3}\approx \mbox{(2.5$\times 10^{12}$ GeV)(1/2--1)},\\
\begin{array}{l}
M_{1} \approx  |x-z^2|M_{3} \sim 10^{10} \mbox{\
GeV}(1/4\mbox{--}2).
\end{array}
\end{array}
\end{eqnarray}

\noindent Note that both the RH neutrinos and the light neutrinos
have hierarchical masses.

In the BPW model of Ref. \cite{BPW}, the parameters $\sigma, \eta,
\epsilon\ etc.$ were chosen to be real. To allow for CP violation,
this framework was extended to include phases for the parameters
in Ref. \cite{BPR}. Remarkably enough, it was found that there
exists a class of fits within the SO(10)/G(224) framework, which
correctly describes not only  (a) fermion masses, (b) CKM mixings
and (c) neutrino oscillations \cite{BPW,JCPKeK}, but also (d) the
observed CP and flavor violations in the K and B systems (see Ref.
\cite{BPR} for the predictions in this regard). A representative
of this class of fits (to be called fit A) is given by \cite{BPR}:
\begin{eqnarray}
\label{eq:fitA}
\begin{array}{l}
\ \ \sigma = 0.109-0.012i, \quad \eta = 0.122-0.0464i, \quad
\epsilon =
-0.103, \quad \eta' = 2.4 \times 10^{-3},\\
\begin{array}{l}
\ \epsilon' = 2.35\times 10^{-4}e^{i(69^{\circ})},\quad
\zeta_{22}^d = 9.8\times 10^{-3}e^{-i(149^{\circ})},\quad({\cal
M}_u^0,\ {\cal M}^0_d)\approx (100,\ 1.1) \mbox{ GeV}.
\end{array}
\end{array}
\end{eqnarray}

\noindent In this particular fit $\zeta_{22}^u$ is set to zero for
the sake of economy in parameters. However, allowing for
$\zeta_{22}^u\lsim (1/3)(\zeta_{22}^d)$ would still yield the
desired results. Because of the success of this class of fits in
describing correctly all four features (a), (b), (c) and (d)-which
is a non-trivial feature by itself - we will use fit A as a
representative to obtain the mass-insertion parameters
$\hat{\delta}^{ij}_{LL,RR}$, $(\delta^{ij}_{LL})^{RHN}$ and
$\delta^{ij}_{LR,RL}$ in the lepton sector and thereby the
predictions of our model for lepton flavor violation.

The fermion mass matrices $M_u$, $M_d$ and $M_l$ are diagonalized
at the GUT scale $\approx 2\times 10^{16}$ GeV by bi-unitary
transformations:
 \begin{eqnarray}
 \label{eq:xdxu}
M_{u,d,l}^{diag}\ =\ X_L^{(u,d,l)\dagger}M_{u,d,l} X_R^{(u,d,l)}
\end{eqnarray}

\noindent The analytic expressions for the matrices $X^d_{L,R}$
can be found in \cite{BPR}. The corresponding expressions for
$X^l_{L,R}$ can be obtained by letting ($\epsilon, \epsilon'$)$\to
-3(\epsilon, \epsilon'$).

We now discuss the sources of lepton flavor violation in our
model.

\section{Lepton Flavor Violation in the
SO(10)/G(224) Framework}

We assume that flavor-universal soft SUSY-breaking is transmitted
to the SM-sector at a messenger scale M$^*$, where M$_{GUT}< $
M$^{*}\le$ M$_{string}$. This may naturally be realized e.g. in
models of mSUGRA \cite{mSUGRA}, or gaugino-mediation
\cite{gauginomed}. With the assumption of extreme universality as
in CMSSM, supersymmetry introduces five parameters at the scale
M$^{*}$:
\begin{center}
$m_o, m_{1/2}, A_o, \tan\beta\ {\rm and}\ sgn(\mu).$
\end{center}

\noindent For most purposes, we will adopt this restricted version
of SUSY breaking with the added restriction that $A_o$ = 0 at
M$^{*}$ \cite{gauginomed}. However, we will not insist on strict
Higgs-squark-slepton mass universality. Even though we have flavor
preservation at M$^{*}$, flavor violating scalar
(mass)$^2$--transitions arise in the model through RG running from
M$^*$ to the EW scale. As described below, we thereby have {\it
three sources} of lepton flavor violation.

\noindent {\bf (1) RG Running of Scalar Masses from M$^{*}$ to
M$_{\rm GUT}$.}

With family universality at the scale M$^{*}$, all sleptons have
the mass m$_o$ at this scale and the scalar (mass)$^2$ matrices
are diagonal. Due to flavor dependent Yukawa couplings, with $h_t
= h_b = h_\tau (= h_{33})$ being the largest, RG running from
M$^{*}$ to M$_{\rm GUT}$ renders the third family lighter than the
first two (see e.g. \cite{BHS}) by the amount:
\begin{eqnarray}
\label{eq:deltambr} \Delta\hat{m}_{\tilde{b}_{L}}^2 =
\Delta\hat{m}_{\tilde{b}_{R}}^2 =
\Delta\hat{m}_{\tilde{\tau}_{L}}^2 =
\Delta\hat{m}_{\tilde{\tau}_{R}}^2 \equiv\Delta\approx
-\bigl(\frac{30m_o^2}{16\pi^2}\bigr) h_t^2\ ln(M^{*}/M_{GUT})~.
\end{eqnarray}
The factor 30$\to$12 for the case of G(224). The slepton
(mass)$^2$ matrix thus has the form $\tilde{\rm
M}^{(o)}_{\tilde{l}}$ = diag(m$_o^2$, m$_o^2$, m$_o^2 -\Delta$).
As mentioned earlier, the spin-1/2 lepton mass matrix is
diagonalized at the GUT scale by the matrices X$_{L,R}^l$.
Applying the same transformation to the slepton (mass)$^2$ matrix
(which is defined in the gauge basis), i.e. by evaluating
X$_L^{l\dagger}$($\tilde{\rm M}^{(o)}_{\tilde{l}}$)$_{LL}$ X$_L$
and similarly for L$\to$R, the transformed slepton (mass)$^2$
matrix is no longer diagonal. The presence of these off-diagonal
elements (at the GUT-scale) given by:
\begin{eqnarray}
\label{eq:deltahat} (\hat{\delta}_{LL,RR}^l)_{ij}=
\left(X_{L,R}^{l\dagger}(\tilde{\rm M}^{(o)}_{\tilde{l}})
X_{L,R}\right)_{ij}/m^2_{\tilde{l}}
\end{eqnarray}
\noindent induces flavor violating transitions
$\tilde{l}_{L,R}^i\to \tilde{l}_{L,R}^j$. Here m$_{\tilde{l}}$
denotes an average slepton mass and the hat signifies GUT-scale
values.

\noindent {\bf (2) RG Running of the $A-$parameters from M$^{*}$
to M$_{\rm GUT}$.}

Even if  $A_o$ = 0 at the scale M$^{*}$ (as we assume for
concreteness, see also \cite{gauginomed}). RG running from M$^{*}$
to M$_{\rm GUT}$ induces $A-$parameters at M$_{\rm GUT}$, invoving
the SO(10)/G(224) gauginos; these yield chirality flipping
transitions ($\tilde{l}^i_{L,R}\to \tilde{l}^j_{R,L}$).

Evaluated at the GUT scale, the $A-$parameters, induced
respectively through the couplings $h_{ij}$, $a_{ij}$ and
$g_{ij}$, are given by:
\begin{eqnarray}
\label{eq:RGEA}
\begin{array}{l}
\ \ A_{ij}^{(1)}=\frac{63}{2}\frac{1}{8\pi^2} h_{ij} g_{10}^2
M_{\lambda}
ln(\frac{M^{*}}{ M_{{\bf 10_H}}})\quad ~~~~~~(i,j =2,3)\\
\begin{array}{l}
\ A_{ij}^{(2)}=\frac{95}{2}\frac{1}{8\pi^2}\frac{a_{ij}\langle
{\bf 45_H}\rangle}{M'} g_{10}^2 M_{\lambda}
ln(\frac{M^{*}}{M_{{\bf 10_H}}})\quad (ij =23,12) \\
\begin{array}{l}
A_{ij}^{(3)}=\frac{90}{2}\frac{1}{8\pi^2}\frac{g_{ij}\langle {\bf
10_H}\rangle}{M''} g_{10}^2 M_{\lambda} ln(\frac{M^{*}}{M_{{\bf
16_H}}})\quad (ij =23,22,12)
\end{array}
\end{array}
\end{array}
\end{eqnarray}

The coefficients ($\frac{63}{2},\frac{95}{2},\frac{90}{2}$) are
the sums of the Casimirs of the SO(10) representations of the
chiral superfields involved in the diagrams. For the case of
G(224),
($\frac{63}{2},\frac{95}{2},\frac{90}{2}$)$\to$($\frac{27}{2},\frac{43}{2},\frac{42}{2}$).
Thus, summing $A^{(1)}$, $A^{(2)}$ and $A^{(3)}$, the induced $A$
matrix for the leptons is given by:
\begin{eqnarray}
\label{eq:Al} \small{A^l_{LR} =Z\left\{K_{\bf 10}\left[
\begin{array}{ccc}
0&-285\epsilon'&0\\285\epsilon' &63\zeta_{22}^u&-
285\epsilon+63\sigma\\0& 285\epsilon+63\sigma&63
\end{array}\right] + K_{\bf 16}\left[
\begin{array}{ccc}
0&90\eta'&0\\ 90\eta' &90(\zeta_{22}^d-\zeta_{22}^u)&
90(\eta-\sigma)\\0& 90(\eta-\sigma)&0
\end{array}\right]\right\}}
\end{eqnarray}
\noindent where $Z = \bigl(\frac{1}{16\pi^2}\bigr) h_t g_{10}^2
M_{\lambda}$, $K_{\bf 10}=ln(\frac{M^{*}}{ M_{{\bf 10_H}}})$ and
$K_{\bf 16}=ln(\frac{M^{*}}{ M_{{\bf 16_H}}})$. For simplicity if
we let $M_{{\bf 16_H}}\approx M_{{\bf 10_H}}\approx M_{{\rm
GUT}}$, we can write the $A$ matrix in the SUSY basis as:
\begin{eqnarray}
\label{eq:Ad} A^l_{LR} =Z\ ln(\frac{M^{*}}{M_{{\rm GUT}}})
(X_L^l)^{\dagger}\left[
\begin{array}{ccc}
0&-285\epsilon'+90\eta' &0\\285\epsilon'+90\eta'
&90\zeta_{22}^d-27\zeta_{22}^u&-285\epsilon+90\eta-27\sigma\\0&
285\epsilon+90\eta-27\sigma&63
\end{array}\right] X_R^l
\end{eqnarray}
Approximate analytic forms for X$^d_{L,R}$ are given in Ref.
\cite{BPR}, and X$^l_{L,R}$ can be obtained from X$^d_{L,R}$ by
the substitutions $(\epsilon,\ \epsilon')\to -3(\epsilon,\
\epsilon')$. The chirality flipping transition angles are defined
as :
\begin{eqnarray}
\label{eq:deltalr} (\delta^{l}_{LR})_{ij}\ \equiv\
(A^{l}_{LR})_{ij}\ \bigl(\frac{v_d}{m_{\tilde{l}}^2}\bigr)\ =\
(A^{l}_{LR})_{ij}\ \bigl(\frac{v_u}{\tan\beta\
m_{\tilde{l}}^2}\bigr)~.
\end{eqnarray}

\noindent {\bf (3) RG Running of scalar masses from M$_{\rm GUT}$
to the RH neutrino mass scales:}

We work in a basis in which the charged lepton Yukawa matrix Y$_l$
and M$_R^{\nu}$ are diagonal at the GUT scale. The off-diagonal
elements in the Dirac neutrino mass matrix Y$_N$ in this basis
give rise to lepton flavor violating off-diagonal components in
the left handed slepton mass matrix through the RG running of the
scalar masses from M$_{\rm GUT}$ to the RH neutrino mass scales
M$_{R_i}$. The RH neutrinos decouple below M$_{R_i}$. (For RGEs
for MSSM with RH neutrinos see e.g. Refs. \cite{HNTY} and
\cite{HN}.) In the leading log approximation, the off-diagonal
elements in the left-handed slepton (mass)$^2$-matrix, thus
arising, are given by:
\begin{eqnarray}
\label{eq:RHN} (\delta_{LL}^l)_{ij}^{RHN} =
\frac{-(3m_o^2+A_o^2)}{8\pi^2}\sum_{k=1}^{3}
(Y_N)_{ik}(Y_N^{*})_{jk}\ ln(\frac{M_{GUT}}{M_{R_k}})~.
\end{eqnarray}

\noindent The superscript RHN denotes the contribution due to the
presence of the RH neutrinos. We remind the reader that the masses
M$_{R_i}$ of RH neutrinos are well determined within our framework
to within factors of 2 to 4 (see Eq. (\ref{eq:MajM})). The total
LL contribution is thus:
\begin{eqnarray}
\label{eq:deltatot}
(\delta_{LL}^l)_{ij}^{Tot}=(\hat{\delta}_{LL}^l)_{ij}+(\delta_{LL}^l)_{ij}^{RHN}
\end{eqnarray}

Now, most authors including those using SUSY SU(5) with RHN's or
SUSY SO(10) \cite{SU(5)RHN,SO(10)LFV} have cosidered only the
second term $(\delta^l_{LL})^{RHN}$ that arises due to the
right-handed neutrinos. As mentioned in the introduction, however,
the first term $\hat{\delta}^l_{LL}$ and the contribution of the
$A-$term $\delta^l_{LR,RL}$ (Eq. (\ref{eq:deltalr})) are found to
dominate over the second term (as long as $\ln(M^*/M_{GUT})\sim
1$). We obtain our results by including the contributions from all
three sources listed above in Eqs. (\ref{eq:deltahat}),
(\ref{eq:deltalr}) and (\ref{eq:RHN}). They are presented in the
following section.

\section{Results}
The decay rates for the lepton flavor violating processes l$_i\to$
l$_j \gamma\ (i>j)$ are given by:
\begin{eqnarray}
\label{eq:br} \Gamma(l_i^+ \to l_j^+ \gamma) =\frac{e^2
m_{l_i}^3}{16\pi} \left(|A_L^{ji}|^2 + |A_R^{ji}|^2\right)
\end{eqnarray}

\noindent Here $A_L^{ji}$ is the amplitude for $(l_i)^+_L
\rightarrow (l_j)^+ \gamma$ decay, while $A_R^{ji} = A((l_i)^+_R
\rightarrow (l_j)^+ \gamma)$. The amplitudes $A_{L,R}^{ji}$ are
evaluated in the mass insertion approximation using the
$(\delta_{LL}^l)^{Tot},\ \delta_{RR}^l,\ \delta_{LR,RL}^l$
calculated as above. The general expressions for the amplitudes
$A_{L,R}^{ji}$ in one loop can be found in e.g. Refs. \cite{HNTY}
and \cite{HN}. We include the contributions from both chargino and
neutralino loops with or without the $\mu-$term.

We evaluate the amplitudes by first going to a basis in which the
chargino and the neutralino mass matrices are diagonal. Analytic
expressions for this diagonalization can be obtained in the
approximation $|M_2\pm\mu|$ and $|\frac{5}{3} M_1\pm\mu|\gg m_Z$
and $|M_2\mu|> m_W^2 \sin 2\beta$ \cite{GH}. This approximation
holds for all the input values of $(m_o,\ m_{1/2})$ that we
consider.

In Table 1 as well as in Fig. 1, we give the branching ratios of
the processes $\mu\to e\gamma,\ \tau\to\mu\gamma,\ \tau\to
e\gamma$ for the case of SO(10), with some sample choices of
(m$_o$, m$_{1/2}$). For these calculations we set
ln$\left(\frac{M^{*}}{M_{GUT}}\right)$ = 1, i.e. $M^{*}\approx 3
M_{GUT}$, $\tan\beta = 10$, M$_{R_1} = 10^{10}$ GeV, M$_{R_2} =
10^{12}$ GeV and M$_{R_3} = 5\times 10^{14}$ GeV (see Eq.
(\ref{eq:MajM})), and $A_o$( at M$^*$) = 0. The corresponding
values for G(224) are smaller approximately by a factor of 4 to 6
in the rate, provided $\ln(M^*/M_{GUT})$ is the same in both cases
(see comments below Eqs. (\ref{eq:deltambr}) and (\ref{eq:RGEA})).

\vspace*{12pt}

\noindent
\begin{tabular}{|c|c|c|c|c|c|c|}
\hline \rule[-3mm]{0mm}{8mm} ($m_o,\ m_{1/2}$)//$\tan\beta$ &
\multicolumn{2}{c|}{Br($\mu\to e\gamma$)}&\multicolumn{2}{c|} {Br($\tau\to\mu\gamma$)}&\multicolumn{2}{c|} {Br($\tau\to e\gamma$)}\\
\hline
& $\mu>0$& $\mu<0$ & $\mu>0$ &$\mu<0$ & $\mu>0$ &$\mu<0$ \\

\hline

I (600, 300)//10 & 3.3$\times10^{-12}$ & 9.8$\times 10^{-12}$
         & 2.4$\times 10^{-9}$  & 3.1$\times10^{-9}$  & 2.4$\times 10^{-12}$  & 3.3$\times10^{-12}$
                  \\ \hline
II (800, 250)//10 & 2.9$\times10^{-13}$ & 1.7$\times 10^{-12}$
         & 1.9$\times 10^{-9}$  & 1.9$\times10^{-9}$  & 2.0$\times 10^{-12}$  & 2.0$\times10^{-12}$
                  \\\hline
III (450, 300)//10 & 2.7$\times10^{-11}$ & 4.6$\times 10^{-11}$
         & 2.7$\times 10^{-9}$  & 5.6$\times10^{-9}$  & 2.7$\times 10^{-12}$  & 6.1$\times10^{-12}$
                  \\\hline
IV (500, 250)//10 & 5.9$\times10^{-12}$ & 1.9$\times 10^{-11}$
         & 4.8$\times 10^{-9}$  & 6.4$\times10^{-9}$  & 5.0$\times 10^{-12}$  & 6.9$\times10^{-12}$
                  \\\hline
V (100, 440)//10 & 1.02$\times10^{-8}$ & 1.02$\times 10^{-8}$
         & 8.3$\times 10^{-8}$  & 8.4$\times10^{-8}$  & 1.0$\times 10^{-10}$  & 1.0$\times10^{-10}$
          \\ \hline
VI (1000, 250)//10 & 1.6$\times10^{-13}$ & 5.6$\times 10^{-12}$
         & 9.5$\times 10^{-10}$  & 9.0$\times10^{-10}$  & 1.0$\times 10^{-12}$  & 9.5$\times10^{-13}$
          \\ \hline
VII (400, 300)//20 & 9.5$\times10^{-12}$ & 3.8$\times 10^{-11}$
         & 1.4$\times 10^{-8}$  & 1.8$\times10^{-8}$  & 1.5$\times 10^{-11}$  & 1.9$\times10^{-11}$
                  \\\hline
\end{tabular}

\vskip.10in Table 1. {\small Branching ratios of $l_i\to
l_j\gamma$ for the SO(10) framework with $\kappa\equiv
\ln(M^*/M_{GUT})=1$; ($m_o,\ m_{1/2}$) are given in GeV, which
determine $\mu$ through radiative electroweak symmetry breaking
conditions. The entries for Br($\mu\to e\gamma$) for the case of
G(224) would be reduced by a factor $\approx 4-6$ compared to that
of SO(10) (see text). } \vspace*{12pt}

To give the reader an idea of the magnitudes of the various
contributions, we exhibit in table 2 the amplitudes for the
process $\mu\to e\gamma$ calculated individually from the four
sources $\hat\delta^{ji}_{LL},\ \delta^{ji}_{LR,RL}$ and
$(\delta^{ji}_{LL})^{RHN}$ (see Eqs. (\ref{eq:deltahat}),
(\ref{eq:deltalr}) and (\ref{eq:RHN})), for a few cases of table
1.

\vspace*{12pt}

\noindent
\begin{tabular}{|c|c|c|c|c|}
\hline \rule[-3mm]{0mm}{8mm}
$(m_o,~m_{1/2})$(GeV)&$A_L^{(1)}(\hat{\delta}_{LL}) $&
  $A_L^{(2)}(\delta_{LR})$&
  $A_R(\delta_{RL})$&$A_L^{(3)}((\delta_{LL})^{RHN})$\\
 \hline

I, $(600,~300)$ &$3.3\times 10^{-13}$  & $-6.7\times 10^{-13}$  &
$-5.9\times 10^{-13}$&$2.4\times 10^{-14}$\\ \hline

II, $(800,~250)$ &$2.9\times 10^{-13}$  & $-1.8\times 10^{-13}$ &
$-1.6\times 10^{-13}$&$2.0\times 10^{-14}$  \\ \hline

IV, $(500,~250)$ & $4.8\times 10^{-13}$  & $-9.7\times 10^{-13}$
& $-8.5\times 10^{-13}$&$3.4\times 10^{-14}$\\ \hline

\end{tabular}
\vskip.10in \noindent Table 2. {\small Comparison of the various
contributions to the amplitude for $\mu\to e\gamma$ for cases I,
II and IV, with $\mu>0$. Each entry should be multiplied by a
common factor a$_o$. Imaginary parts being small are not shown.
Note that columns 2,3 and 4 arising from RG running from $M^*\to
M_{GUT}$ (see text) dominate over the RHN contribution.}
\vspace*{5pt}

Glancing at tables 1 and 2, the following features of our results
are worth noting:

\noindent {\bf(1)} It is apparent from table 2 that the
contribution due to the presence of the RH neutrinos\footnote{In
the context of contributions due to the RH neutrinos alone, there
exists an important distinction (partially observed by Barr, see
Ref. \cite{SO(10)LFV}) between the hierarchical BPW form
\cite{BPW} and the lop-sided Albright-Barr (AB) form \cite{AB} of
the mass-matrices. The amplitude for $\mu\to e\gamma$ from this
source turns out to be proportional to the difference between the
(23)-elements of the Dirac mass-matrices of the charged leptons
and the neutrinos, with (33)-element being 1. This difference is
(see Eq. (\ref{eq:mat})) is $\eta-\sigma\approx 0.041$, {\it which
is naturally small for the hierarchical BPW model} (incidentally
it is also $V_{cb}$), while it is order one for the lop-sided AB
model. This means that the rate for $\mu\to e\gamma$ due to RH
neutrinos would be about $600$ times larger in the AB model than
the BPW model (for the same input SUSY parameters). } (fifth
column) is about an order of magnitude smaller, {\it in the
amplitude}, than those of the others (proportional to
$\hat{\delta}^{ij}_{LL},\ \delta^{ij}_{LR}$ and
$\delta^{ij}_{RL}$), listed in columns 2, 3 and 4. The latter
arise from RG running of the scalar masses and the $A-$parameters
in the context of SO(10) or G(224) from $M^*$ to M$_{GUT}$. It
seems to us that the latter, which have commonly been omitted in
the literature, should exist in any SUSY GUT model for which the
messenger scale for SUSY-breaking is high ($M^*>M_{GUT}$), as in a
mSUGRA model. {\it The inclusion of these new contributions to LFV
processes arising from post-GUT physics, that too in the context
of a predictive and realistic framework, is the distinguishing
feature of the present work.}

\noindent {\bf(2)} Again from table 2 we see that the two dominant
contributions to $A_L = A(\mu^+_L\to e^+\gamma)$, arising from
$\delta_{LL}$ and $\delta_{LR}$-insertions, partially cancel each
other if $\mu>0$; they would however add if $\mu<0$. By contrast,
$A_R$ gets contribution dominantly only from $\delta_{RL}$ (column
4).\footnote{Although $\hat{\delta}_{RR}$ is comparable to
$\hat{\delta}_{LL}$, its contribution to $A_R$ (via the bino loop)
is typically suppressed compared to that of $\delta_{LL}$ to $A_L$
(in part by the factor $(\alpha_1/\alpha_2)(M_1/M_2)$) in most of
the parameter space.} As a result we find that in our model,
typically, $|A_R|>|A_L|$ if $\mu>0$ and $|A_L|>|A_R|$ if $\mu<0$.

\noindent {\bf(3)} Owing to the general prominence of the new
contributions from post-GUT physics, we see from table 1 that case
V, (with low $m_o$ and high $m_{1/2}$) is clearly excluded by the
empirical limit on $\mu\to e \gamma$-rate (see Sec. 1). Case III
is also excluded, for the case of SO(10), yielding a rate that
exceeds the limit by a factor of about 2 (for
$\kappa=\ln(M^*/M_{GUT})\gsim 1$), though we note that for the
case of G(224), Case III is still perfectly compatible with the
observed limit (see remark below table 1). All the other cases (I,
II, IV, VI, and VII), with medium heavy ($\sim$ 500 GeV) to
moderately heavy sleptons (800-1000 GeV), are compatible with the
empirical limit, even for the case of SO(10). The interesting
point about these predictions of our model, however, is that
$\mu\to e \gamma$ should be discovered, even with moderately heavy
sleptons, both for SO(10) and G(224), with improvement in the
current limit by a factor of 10--100. Such an improvement is being
planned at the forthcoming MEG experiment at PSI.

\noindent {\bf(4)} We see from table 1 that $\tau\to\mu\gamma$
(leaving aside case V, which is excluded by the limit on $\mu\to
e\gamma$), is expected to have a branching ratio in the range of
$2\times 10^{-8}$ (Case VII) to about $(1\ {\rm or}\  2)\times
10^{-9}$ (Case VI or II). The former may be probed at BABAR and
BELLE, while the latter can be reached at the LHC or a super B
factory. The process $\tau\to e\gamma$ would, however, be
inaccessible in the foreseeable future (in the context of our
model).

\noindent {\bf(5)} {\bf The WMAP-Constraint:} Of the cases
exhibited in table 1, Case V ($m_o = 100$ GeV, $m_{1/2} = 440$
GeV) would be compatible with the WMAP-constraint on relic dark
matter density, in the context of CMSSM, assuming that the
lightest neutralino is the LSP and represents cold dark matter
(CDM), accompanying co-annihilation mechanism. (See e.g.
\cite{JEllis}). As mentioned above (see table 1), a spectrum like
Case V, with low $m_o$ and higher $m_{1/2}$, is however excluded
in our model by the empirical limit on $\mu\to e\gamma$. {\it Thus
we infer that in the context of our model CDM cannot be associated
with the co-annihilation mechanism.}

Several authors (see e.g. Refs.\cite{Olive} and \cite{Baer}),
have, however considered the possibility that Higgs-squark-slepton
mass universality need not hold even if family universality does.
In the context of such non-universal Higgs mass (NUHM) models, the
authors of Ref. \cite{Baer} show that agreement with the WMAP data
can be obtained over a wide range of mSUGRA parameters. In
particular, such agreement is obtained for ($m_\phi/m_o$) of order
unity (with either sign) for almost all the cases (I, II, III, IV,
VI and VII)\footnote{We thank A. Mustafayev and H. Baer for
private communications in this regard.}, with the LSP (neutralino)
representing CDM.\footnote{We mention in passing that there may
also be other posibilities for the CDM if we allow for either
non-universal gaugino masses, or axino or gravitino as the LSP, or
R-parity violation (with e.g. axion as the CDM). } (Here
$m_\phi\equiv sign(m^2_{H_{u,d}})\sqrt{|m^2_{H_{u,d}}|}$, see
\cite{Baer}). All these cases (including Case III for G(224)) are
of course compatible with the limit on $\mu\to e\gamma$.

\noindent {\bf(6)} {\bf Coherent $\mu - e$ conversion in nuclei:}
In our framework, $\mu-e$ conversion (i.e. $\mu^- + N \rightarrow
e^- + N$)
 will occur when the photon emitted in the virtual decay
$\mu \rightarrow e \gamma^*$ is absorbed by the nucleus (see e.g.
 \cite{marciano}). In such situations, there is a rather simple relation connecting
the $\mu-e$ conversion rate with $B(\mu \rightarrow e \gamma)$:
$B(\mu \rightarrow e \gamma)/
(\omega_{conversion}/\omega_{capture}) = R \simeq (230-400)$,
depending on the nucleus.  For example, $R$ has been calculated to
be $R \simeq 389$ for $^{27}Al$, 238 for $^{48}Ti$ and 342 for
$^{208}Pb$ in this type of models. (These numbers were computed in
\cite{marciano} for the specific model of \cite{BHS}, but they
should approximately hold for our model as well.) With the
branching ratios listed in Table 1 ($\sim 10^{-11}$ to $10^{-13}$)
for our model, $\omega_{conversion}/\omega_{capture} \simeq$
(40--1) $\times 10^{-15}$. The MECO experiment at Brookhaven is
expected to have a sensitivity of $10^{-16}$ for this process, and
thus will test our model.

\noindent {\bf(7)} {\bf Parity odd asymmetry in $\mu^+ \rightarrow
e^+ \gamma$ decay:} Parity violation can be observed by studying
the correlation between the momentum $\vec{p_e}$ of $e^{+}$ in
$\mu^+ \rightarrow e^+ \gamma$ decay and the polarization vector
$\vec{P}$ of positive muons (from $\pi^+$ decays). The
distribution of $e^+$ is proportional to
($1+\mathcal{A}~\hat{p}_e.\vec{P}$) where $\mathcal{A}$ is the
$P$--odd asymmetry parameter given by $\mathcal{A}(\mu^+
\rightarrow e^+ \gamma) = (|A_L|^2 - |A_R|^2 )/ (|A_L|^2 +
|A_R|^2)$. Here $A_L$ is the amplitude for $\mu^+_L \rightarrow
e^+ \gamma$ decay, while $A_R = A(\mu^+_R \rightarrow e^+
\gamma$). In our model, as noted in (2), we typically have
$|A_R|>|A_L|$ and thus $\mathcal{A}(\mu^+ \rightarrow e^+
\gamma)<0$ if $\mu>0$, and $|A_L|>|A_R|$ and thus $\mathcal{A}>0$
if $\mu<0$. For example, with $(m_o, m_{1/2}) = (800, ~250)$ GeV,
$\mu>0$ and $\tan\beta = 10$, we obtain
$|A_L|=|A_L^{(1)}(\hat{\delta}_{LL})+A_L^{(2)}(\delta_{LR})+A_L^{(3)}|=1.3\times
10^{-13}$ (see table 2) while $|A_R| \simeq 1.6 \times 10^{-13}$,
and thus $\mathcal{A}\simeq -0.25$, while for $(m_o, m_{1/2})
=(500,~250)$ GeV and $\tan\beta = 10$ we get, $|A_L| \simeq 4.7
\times 10^{-13}$ and $|A_R| \simeq 8.6 \times 10^{-13}$, yielding
$\mathcal{A} \simeq -0.54$. The precise prediction of our model
for ${\mathcal A}$ would thus be definitive once the SUSY spectrum
is known.

We can compare the predictions of our model for $\mathcal{A}$ with
those of other SUSY models. In the MSSM with $\nu_R$, since LFV
arises through $\delta_{LL}$ type mixings, $A_L \gg A_R$, and thus
$\mathcal{A}(\mu^+ \rightarrow e^+ \gamma$) $\approx +1$, at least
for $\tan\beta \leq 30$ or so, regardless of the choice of $(m_o,
m_{1/2})$. In SUSY $SU(5)$ GUT, with or without $\nu_R$, the GUT
threshold effects realized in the regime $M_{GUT} \leq \mu \leq
M_*$ generate $\delta_{RR}$ type mixings, and will lead to $A_R
\gg A_L$ and thus $\mathcal{A} \simeq -1$. In the SUSY $SO(10)$
models with symmetric mass matrices, such as the ones studied in
\cite{BHS,rnm}, $A_L = A_R$ from GUT threshold effects, leading to
a vanishing $\mathcal{A}$. Thus, we see that a determination of
$\mathcal{A}$ may help sort out the specific type of GUT that is
responsible for LFV.

\noindent {\bf(8)} {\bf Correlation between muon $g-2$ and $\mu
\rightarrow e \gamma$}: Currently there exists a discrepancy
between theory and experiment in the anomalous magnetic moment of
the muon:  $\Delta a_\mu = a_\mu^{\rm expt} - a_\mu^{SM} = 251(93)
\times 10^{-11}$ \cite{marciano1}.  This is a 2.7 sigma effect
\footnote{This analysis is based on theory and data on $e^+ e^-
\rightarrow$ hadron. If $\tau\to \nu_\tau +$ hadron data is used,
this discrepancy reduces to 1.3 sigma; this may however be less
reliable \cite{marciano1}.} and may be an indication of low energy
supersymmetry.  In our framework, this discrepancy can be
considerably reduced for some, but not all, choices of the SUSY
spectrum. When the sleptons are relatively light ($\leq 500$ GeV)
with $\tan\beta = 10-20$, the SUSY contribution to $a_\mu$ is in
the range $(50 - 200) \times 10^{-11}$.  For example, following a
recent numerical analysis (see \cite{baer} and references there
in), we find $\Delta a_\mu^{SUSY} \approx 180\times 10^{-11}$ for
the cases of both IV and VII (see table 1). Note that when the
SUSY contributions to $\Delta a_\mu$ becomes significant, $B(\mu
\rightarrow e \gamma)$ is enhanced.  Thus, a confirmation of new
physics contribution to $a_\mu$, for example by improved precision
in the $e^+ e^- \rightarrow $ hadron data and in the theoretical
analysis, would imply (in the context of a SUSY-explanation) that
$\mu \rightarrow e \gamma$ is just around the corner, within our
framework.

In summary, lepton flavor violation is studied here within a
predictive SO(10)/G(224)-framework, possessing supersymmetry, that
was proposed in Refs. \cite{BPW,BPR}. The framework seems most
realistic in that it successfully describes five phenomena: (i)
fermion masses and mixings, (ii) neutrino oscillations, (iii) CP
violation, (iv) quark flavor-violations, as well as (v)
baryogenesis via leptogenesis \cite{PatiLepto}. LFV emerges as an
important prediction of this framework bringing no new parameters,
barring the few flavor-preserving SUSY parameters.

As mentioned before, the inclusion of contributions to LFV arising
both from the presence of the RH neutrinos as well as those from
the post-GUT regime, that too within a realistic framework, is the
distinguishing feature of the present work. Typically, the latter
contribution, which is commonly omitted in the literature, is
found to dominate. Our results show that -- (i) The decay $\mu\to
e\gamma$ should be seen with improvement in the current limit by a
factor of 10 -- 100, even if sleptons are moderately heavy ($\sim
800$ GeV, say); (ii) for the same reason, $\mu-e$ conversion ($\mu
N \to e N$) should show in the planned MECO experiment, and (iii)
$\tau\to\mu\gamma$ may be accessible at the LHC and a super
B-factory. It is noted that the muon ($g-2$)-anomaly, if
confirmed, would strongly suggest, within our model, that the
discovery of the $\mu\to e\gamma$ decay is imminent. The
significance of a measurement of the parity-odd asymmetry in
polarized $\mu^+$ decay into $e^+\gamma$ is also noted. In
conclusion, the SO(10)/G(224) framework pursued here seems most
successful on several fronts; it can surely meet further stringent
tests through a search for lepton flavor violation.

\section{Acknowledgements}
We would like to thank Howard Baer, Azar Mustafayev, Keith Olive,
Serguey Petcov and Yasutaka Takanishi for helpful correspondence.
The work of KSB is supported in part by the Department of Energy
Grant Nos. DE-FG02-04ER46140 and DE-FG02-04ER41306. That of JCP is
supported in part by Department of Energy Grant No.
DE-FG-02-96ER-41015.

\begin{figure}[!ht]
\centering
\includegraphics[scale=5,height=3in,width=5in]{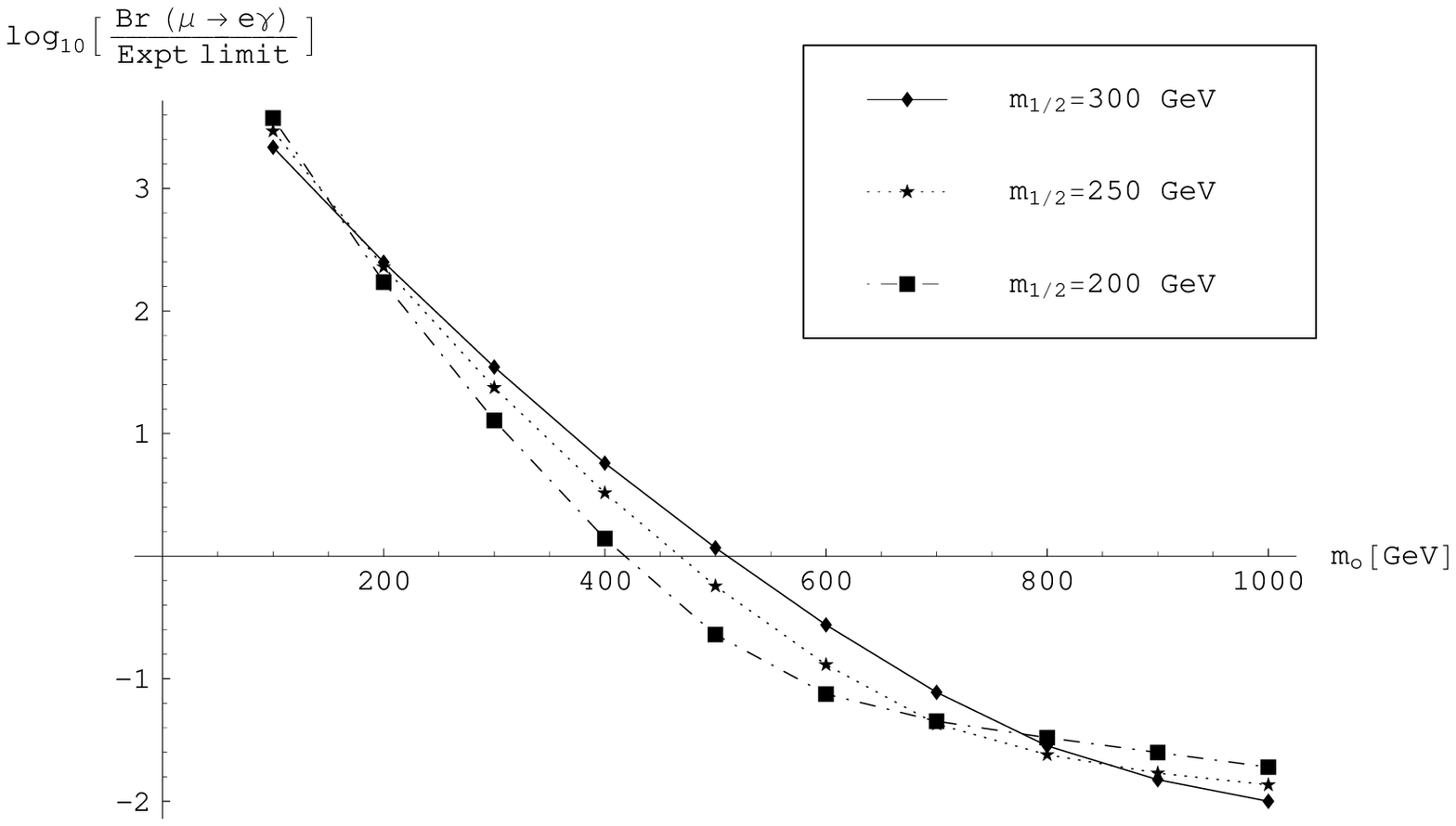}
\end{figure}

\vskip.10in Fig. 1. {\small Log of Br($\mu\to e\gamma$) divided by
the experimental bound ($1.2\times 10^{-11}$) obtained for the
SO(10) framework with $\ln(M^*/M_{GUT})=1$, $\tan\beta=10$ and
$\mu>0$ vs $m_o$ (in GeV) with $m_{1/2}=200,~250$ and $300$ GeV.}
\vspace*{12pt}

\end{document}